# Deposition of defected graphene on (001) Si substrates by thermal decomposition of acetone

T.I. Milenov[1], I. Avramova[2], E. Valcheva[3], G.V. Avdeev[4], S. Rusev[3], S. Kolev[1], I. Balchev[1], I. Petrov[1], D. Pishinkov[1] and V.N. Popov[3]

[1]"E. Djakov" Institute of Electronics, Bulgarian Academy of Sciences, 72 Tzarigradsko Chaussee Blvd., 1784 Sofia, Bulgaria

[2]Institute of General and Inorganic Chemistry, Bulgarian Academy of Sciences, Acad. G. Bonchev Str., bl. 11, 1113 Sofia, Bulgaria

[3]Faculty of Physics, University of Sofia, 5 James Boucher Blvd., 1164 Sofia, Bulgaria

[4]"R. Kaishev" Institute of Physical Chemistry, Bulgarian Academy of Sciences, Acad. G. Bonchev Str., bl. 11, 1113 Sofia, Bulgaria

Abstract

We present results on the deposition and characterization of defected graphene by the chemical vapor deposition (CVD) method. The source of carbon/carbon-containing radicals is thermally decomposed acetone ($C_2H_6CO$) in Ar main gas flow. The deposition takes place on (001) Si substrates at about 1150-1160$^0$C. We established by Raman spectroscopy the presence of single- to few- layered defected graphene deposited on two types of interlayers that possess different surface morphology and consisted of mixed $sp^2$ and $sp^3$ hybridized carbon. The study of interlayers by XPS, XRD, GIXRD and SEM identifies different phase composition: i) a diamond-like carbon dominated film consisting some residual SiC, $SiO_2$ etc.; ii) a $sp^2$- dominated film consisting small quantities of $C_{60}/C_{70}$ fullerenes and residual Si-O-, C=O etc. species. The polarized Raman studies confirm the presence of many single-layered defected graphene areas that are larger than few microns in size on the predominantly amorphous carbon interlayers.

## 1.Introduction

Graphene is one-atom thick layered material, which consists of completely $sp^2$-bonded carbon atoms, tightly-packed into a honeycomb lattice. It has a number of unique properties like low optical absorption, high electrical conductivity, high mechanical strength, interesting membrane properties, etc. Several different paths for synthesizing graphene have been followed experimentally during the last decade. However, only the thermally- and plasma-assisted (PA) chemical vapor deposition (CVD) on metal substrates (copper, nickel, etc.) (Kim et al. [1], Reina et al. [2], etc.) and epitaxial growth on SiC substrates (Berger et al. [3,4]) have been developed industrially. The latter method is based on C (or Si) termination of the $(0001)_C$ (or $(0001)_{Si}$) SiC surface and requires high temperature and expensive SiC substrates. The former method is based on PA thermal decomposition of a carbon-containing precursor on a catalytic metal surface. This method provides high reliability and relatively high quality of the graphene films and presently there are many suppliers of reactors for PACVD of graphene. The most preferred precursor for this method is methane ($CH_4$) because the chemical bond in $CH_4$ is relatively strong and prevents fast decomposition of the reagent at temperatures below $1000^0C$ (Muñoz and Gómez-Aleixandre [5]). The main challenge in the application of the PACVD method is that even if the synthesized graphene was perfect, it cannot be used directly in microelectronics and has to be transferred from metal to another surface, which usually generates a large number of defects. Therefore, the problem with the deposition of graphene on silicon or surfaces compatible with silicon technology is still unsolved. We investigated the possible application of acetone as a precursor in a thermally-assisted CVD and showed in a recently published work (Milenov et al. [6]) that few-layered defected graphene/folded graphene can be deposited on commercially available metal foils like Ni, ($Cu_{0.5}Ni_{0.5}$), μ-metal and stainless steel SS 304. Here, we are aim at studying the possible deposition of graphene/graphene related phases (few-layered defected graphene, polygraphene etc.) on (001) Si substrates via thermal decomposition of acetone.

**2. Experimental**

*2.1 CVD process*

We used 2 inch (001) Si substrates and a horizontal-tube quartz CVD reactor with internal diameter of about 70 mm. The experimental setup also consists of a gas-supply system (inlet and outlet parts), a barboteur for evaporation of acetone which is situated in a thermostat, a quartz substrate support and a resistive heating furnace (Fig. 1). The CVD process is based on thermal decomposition of acetone in Ar main gas flow. The temperature of the thermostat was kept at $12^0C$. The heating of the reactor from room temperature to working temperature takes place at a rate of about $300^0C/h$ in atmosphere of pure Ar flow of about 120-180 $cm^3$/min. In order to prevent the super-saturation in the high-temperature zone of the reactor we used a "pulsed" regime in some experiments by alternating the flow of the gas mixture of (Ar + $C_3H_6O$) for 3 min on top of the main flow of pure Ar of about 150-180 $cm^3$/min for 1.5 min for each pulse. The parameters of the deposition processes are summarized in Table 1. The reactor with deposited samples was cooled after the deposition at a rate of about 150-200$^0$C/h.

*2.2 Characterization*

We characterized the carbon layers by Raman spectroscopy, optical and scanning electron microscopy (SEM), X-ray Photoelectron Spectroscopy (XPS), X-ray powder diffraction (XRD) and by grazing incidence beam XRD (GIXRD).

The Raman measurements were carried out using a micro-Raman spectrometer HORIBA Jobin Yvon Labram HR 800 Visible with a He-Ne (633 nm) laser. The laser beam with 0.5 mW power was focused on a spot of about 1 μm in diameter on the studied surfaces, the spectral resolution being 0.5 cm$^{-1}$ or better.

We used LYRA TESCAN scanning electron microscope at 30 kV accelerating voltage for the SEM studies without deposition of any conducting amorphous film on the studied specimens.

X-ray photoelectron spectroscopy was applied for the investigation of the graphene thin films grown by CVD over silicon substrate. The XPS spectra were taken from different points of the surface of the film in order to monitor the growth mechanism of the films as well as their thickness uniformity and homogeneities. XPS measurements were performed on a Kratos AXIS Supra spectrometer with a monochromatic Al K$_\alpha$ (1486.6 eV) source in vacuum better than 10$^{-8}$ Pa at 90 degree take-off angle. Each measurement was initiated with a survey scan from 0 to 1200 eV, pass energy of 160 eV at steps of 1 eV with 1 sweep. For the high resolution measurements, the number of sweeps was increased, the passed energy was lowered to 20 eV at steps of 100 meV. The C1s photoelectron line at 284.6 eV was used for calibration of spectra. The surface composition was determined from the ratio of the corresponding peak intensities, corrected with the photo ionization cross sections. The C1s, O1s, N1s, Si2p photoelectron lines as well the survey scan have been recorded. The point-focused XPS measurements were performed in series of five points on a linear segment and separated by 0.5 mm for each characterized sample.

XRD measurements were performed using PANalytical Empyrean apparatus in different geometries (φ, (θ- 2θ) and (ω- 2θ) scans). The most prominent features in the XRD patterns of the initial (θ-2θ) scans are reflections at 2θ= 26.24$^0$ and 69.07$^0$, which correspond to d$_{(002)}$ of graphitic carbon (ICSD -31170) and $d_{(004)}$ of Si substrate (ICSD-29287). Further on, φ scans (in the range φ= 0- 360$^0$) with reference to $C_{(002)}$ and $Si_{(004)}$ reflections were performed. Finally, GIXRD measurements (in ω-2θ scans) were conducted at ω=0.10$^0$ and ω=0.30$^0$ because the critical angle for total reflection of graphene is 0.21$^0$ (T. Schumann et al. [7]) and the penetration depth at grazing angle of 0.10$^0$ in carbon does not exceed 20-25 Å.

**3. Results**

*3.1 Optical and scanning electron microscopy*

Two areas with different surface morphology were observed by optical microscopy, as shown in Fig. 2. Namely, a clear relief of ridge-like formations, lying along <001>, covers the central area denoted as "R", while the surrounding area, denoted as "H", is covered by an optically inhomogeneous film with a constant thickness. It should be noted that some optical inhomogeneity is also observed on the R areas.

The SEM observations (Figs. 3 a-f) show that the surface relief on the R areas is clearly distinguishable in both secondary electron (SE) and backscattered electron (BSE) regimes. Moreover, a lot of sub-micron crystallites in different grades of grey are observed in the BSE images (Fig. 3 d) indicating a different chemical composition of the crystallites situated on the ridge-like relief on the R areas. On the other hand, the observed contrast changes of the SE image of the H area with submicron dimensions (in light-grey color in Fig. 3 e) should be related to carbon flakes because the corresponding BSE image (Fig. 3 f) does not show any change of the contrast. It should be noted that the surface morphology of the thin films deposited in different experimental series, revealed by optical (Fig. 2) and scanning electron microscopies (Fig. 3 a- f), is completely the same.

*3.2 XPS studies: point XPS study of irradiated area of diameter of 650 μm*

It is clearly established that the C1s photoelectron lines provide information about the existence of $sp^2$- and $sp^3$-hybridized carbon bonds, as well as other types of bonds on the surface of the films within about 5 to 10 nm depth. In Fig. 4, we present the results of the XPS study, conducted in 5 neighboring points on the surface of a sample from the 3PTH experimental series. The analyzed points 1- 4 comprise both H and R areas in different ratio, but the R area in point 2 significantly exceeds the R area of points 1, 3 and 4. The results are representative for all XPS measurements of all studied samples.

C1s photoelectron spectra, taken from all five examined points, are shown in Fig.4a and a rescaled one is shown in Fig. 4b. The nonsymmetrical C1s photoelectron peaks are situated at 284.4 eV (Fig. 4a) and are broadened towards higher energies (Fig. 4b), which points to co-existence of $sp^2$- and $sp^3$-hybridized carbon. Analysis of the higher binding-energy side (296-300 eV) reveals characteristic features of different types of fullerenes (clearly distinguishable shake-up satellites of the C1s peak, Fig. 4b). The shake-up satellites in this region are ascribed to the unchanged electronic density of π and σ orbitals, when planar graphene is curved to form fullerene. Then the difference in the shake-up structure is that the plasmon is a solid state phenomenon in graphite, while it is a collective oscillation of the charge of the single molecule in the fullerene/s (Ref. [8]). According to this work, these shake-ups appear at different energies depending on their origin: the graphite shake-up appears at 290.9 eV, those of $C_{84}$ at 288eV and 291 eV, while $C_{70}$ and $C_{60}$ show shake-ups at 287.5 , 289, 291 eV and 287, 290, 291, 294, 298 eV, respectively. The most pronounced features in the binding energies interval 296- 300 eV (Fig. 4 b), are features at 287.5 (shoulder), 289 and 291 eV. Several weak peaks at 287 (very weak shoulder), 290 (in points 2, 3 and 4), and 294 eV are also distinguishable. The $C_{60}$ shake-up at 298 (not shown in Fig. 4 b) is observable in the spectrum taken from points 4 and 2. In addition it should be noted that the shake-up satellite of graphene appears at 290 eV and overlaps with those of graphite and $C_{60}$ ( Ref. [9]).

The C1s peaks, measured in different points, were subjected to fitting procedure for calculation of the ratio of $sp^2$- to $sp^3$-hybridized carbon and evaluation of the presence of other types of carbon bonds [10-12]. As an example, only fitted C1s spectra, recorded at point 1 and point 2, are shown in the Fig. 5 a and b, respectively. It is worth noticing that, additionally to graphene and fullerene, SiC type bond was evident in the spectra of C1s, taken at the same position (see C1s for point 2). The calculated $sp^2/sp^3$ ratios as well as the concentrations of the constituent elements in the film are summarized in Table 2.

The presence of silicon in Si-O and $SiO_2$ groups was detected from the observed peaks in the Si2p spectra at 101.8 eV and at 103.4 eV (Fig. 6 a and b), respectively [13]. The quantity of Si-O and $SiO_2$ varies at the different positions on the surface. In the case of higher amount of $SiO_2$, we detect also a small amount of SiC (peak at 100.3 eV) in the film (point 2) (Ref. [13]).

From the obtained results, we conclude that graphene film is deposited on the Si substrates. The presence of different interlayers between graphene and the Si substrate can also be deduced. The quantity of $sp^3$-hybridized carbon is significantly higher in the R areas compared to the H areas and, therefore, it can related to diamond-like carbon (DLC). On the contrary, the H area is dominated by $sp^2$-hybridized carbon and can be related to amorphous carbon (aC). The observed SiC and the increased $SiO_2$ content in the DLC-rich areas allows concluding that parallel deposition of SiC and complete oxidation of Si- substrate takes place during the initial steps of the deposition process. It can be assumed that the remaining part of the surface is also oxidized but via Si-O groups, and further formation of a more complex layer, dominated by $sp^2$-hybridized carbon.

*3.3 XRD studies:*

We used φ scans in the range $\varphi = 0\text{-}360^0$ with reference to $d_{(002)}$ of graphitic carbon at $2\theta = 26.6^0$ (see ICSD -31170) and $d_{(004)}$ of silicon at $2\theta = 69.13^0$ (see ICSD-29287) reflections (Fig. 7a) and in order to determine the most suitable orientation for further XRD studies. The distribution of the diffracted signal from graphitic (002) planes with φ (the blue trace in Fig. 7 a) shows that the orientation of the deposited layers follows the orientation of the substrate because the number of counts per second is doubled at $2\theta \sim 73^0$ and $\sim 253^0$, but this dependence is very weak.

XRD patterns, recorded in (θ- 2θ) scans, shown by the blue traces in Figs. 7 b, c and d, reveal the presence of SiC (ICSD-24217) at $2\theta = 35.7^0$, graphitic carbon (ICSD -31170), several features of $C_{70}$ fullerenes (ICSD- 75506) (marked by 1-4 in Fig. 7 b and c) in the range of $(12\text{-}24)^0$ values of 2θ, and the most prominent feature in Fig. 7 d - the merged reflections of $d_{(023)} = d_{(113)}$ at $2\theta = 17.40^0$ and $d_{(130)}$ at $2\theta = 17.70^0$ (see (ICSD- 75506)). The absence of these features in the (ω-2θ) scans points to the conclusion that fullerenes are formed on the initial stages of the deposition and are situated near to the interface Si substrate/interlayer. This result supports the XPS results and the conclusion that the interlayers contain some $C_{70}$ fullerenes.

The GIXRD measurements of the sample from the 3PHT series (Fig. 7 b) shows a very weak reflection of $d_{(111)}$ of SiC (at $2\theta = 35.75^0$) with increasing intensity for decreasing grazing angle, indicating a possible increased content of SiC in the volume of the film. The GIXRD measurements of the sample from the STHT series (Fig. 7 c panel) reveals both AB- and AA-stacked carbon ($C^{AB}_{(002)}$ at $2\theta= 26.57^0$ corresponding to $d_{(002)}= 3.35$ Å as well as $C^{AA}_{(002)}$ at $2\theta= 26.08^0$ corresponding to $d_{(002)}= 3.53$ Å, respectively, Lee et al. [14]) as well as the weak peak at $2\theta= 35.75^0$ Å, corresponding to SiC. The intensity ratios of the reflections of $C^{AA}_{(002)}$ to $C^{AB}_{(002)}$ decrease significantly with increasing $\omega$, which points to increased relative mass of the AB-stacked graphite with depth related to the AA-stacked few-layered graphene phase, situated on the surface. The GIXRD patterns of the sample from the MTHT series are similar to those from the sample from the STHT series, but we did not observe a clear peak of AA-stacked carbon. Moreover, a weak peak of {111} diamond planes (ICSD- 28857) is observed in the ($\omega$-$2\theta$) scan at $\omega=0.1^0$. This weak reflection disappears with increasing $\omega$ for the ($\theta$- $2\theta$) scans as well and, therefore, it can be concluded that this phase appears on the surface of the deposited layers.

*3.4 Unpolarized and polarized Raman spectroscopy of as-deposited samples*

The Raman spectrum of graphene has a clearly established fingerprint [15]. The main first-order features in the Raman spectra of graphene and defected graphene, excited at 633 nm wavelength, are the following:

-G band (~1582 cm$^{-1}$) is the only band in graphene allowed by selection rules for the first-order Raman scattering. It is ascribed to the optical doubly-degenerate phonons of $E_{2g}$ symmetry at the $\Gamma$ point, initially described by Tuinstra and Koenig [16];

-D band (~1330 cm$^{-1}$) is due to breathing-like motion of hexagonal carbon rings, belonging to the transverse optical branch near the K point, and requires a defect for its activation via intervalley double-resonance Raman processes [17];

-D' band (~ 1615 cm$^{-1}$; defect-induced similarly to the D band) occurs via intravalley double-resonance processes [18];

-D" band (~1145 cm$^{-1}$) results from double-resonant intervalley scattering of LA phonons on defects [19]. The intensity of this band is normally ~100 times lower than that of the D band.

Overtones and combination bands:

-2D band is historically known from graphite and carbon-nanotube related literature as the G' band; appears at ~2648-2665 cm$^{-1}$. It is clearly shown [20- 25] that the shape and width of the 2D band can be used for identification of mono-, bi- and three-layered graphene.

-the overtones of the D'- peak (2D'), combination G* (G*= (iTA+LA) phonons), and (D+D') bands occur around 3230, 2450 and 2930 cm$^{-1}$, respectively [26].

The Raman spectrum (excited at 633 nm laser wavelength) taken from the H area of the 3PHT, STHT and MTHT series (Fig. 8), contains all features typical for graphene, namely, a clearly pronounced 2D band with full width at half maximum (FWHM) of 40-58 cm$^{-1}$, $I_{2D}/I_G$ ratio between 2 and 3.5 and $I_{2D}/I_D$ ratio between 2 and 4. However, the 2D band appears at about 2660- 2668 cm$^{-1}$, i.e., it is blueshifted by about 10-15 cm$^{-1}$ and it is somewhat broadened by 10-12 cm$^{-1}$, when its FWHM is ~40 cm$^{-1}$.

The 2D bands are blueshifted by ~10- 20 cm$^{-1}$ and can typically be deconvoluted into: (a) a single Lorentzian with FWHM of ~40-41 cm$^{-1}$ (Fig. 9 a) and (b) four Lorentzians with FWHM of 25 (±1) cm$^{-1}$ with total width of 45- 50 cm$^{-1}$ (Fig. 9 b). The results of the deconvolution indicate the presence of single- and bi-layered defected graphene, respectively [20-25]. Analyzing the inset of Fig 8, it is worth noting, that the 2D band is symmetrical in the spectrum, taken from the 3PHT series, most probably due to the presence of single layered defected graphene, asymmetrically broadened on the higher-energy side of the spectrum from the SHTH series, most probably due to the presence of bi-layered defected graphene, and asymmetrically broadened on the lower-energy side of the spectrum from the MTHT series, most probably due to the presence of multi-layered defected graphene. Comparing the intensity of these bands to the intensity of two-phonon band of the Si substrate (the feature, marked by $II_{Si}$ in Fig. 8) it can be concluded that the thickness of interlayers is lower in the 3PHT films, while it is highest in the MTHT films. In addition, we did not establish a clear difference between the quality of graphene layers deposited on aC- and DLC-rich interlayers, however, bi- and few-layered areas were more frequently observed on DLC-rich interlayers. The results for predominantly single-layered (SL) and bi-layered (BL) defected graphene, according to the deconvolution of 2D bands, are summarized in Table 3. We established that single- and bi- layered graphene is very often observed in 3PHT samples, bi- and multi- layered defected graphene - in STHT samples, while we did not establish single and bi-layered graphene in MTHT samples.

Casiraghi et al. [27] studied theoretically and confirmed experimentally that different graphene edges exhibit variations in the intensity of the D band ($I_D$) in different scattering geometries and concluded that armchair type graphene edges are expected to contribute to $I_D$. The intensity of the D band exhibits a maximum ($I_{Dmax}$) when the incident light is polarized parallel to the armchair type graphene edge and follows a $\cos^2\psi$ law, when the laser light propagates perpendicularly to the graphene layer and the polarization is rotated around the propagation direction by an arbitrary angle ψ. The zigzag graphene edge does not contribute to the intensity of the D band and in Ref. [27] an analytical expression for distinguishing of the different complex graphene edges was derived. We perform similar measurements in $Z(Y_\psi Y_\psi)Z$ as well as in $Z(Y_\psi X_\psi)Z$ geometries, assuming that Z axis is perpendicular to the graphene layer and ψ is the angle between the polarization and the [010] direction of Si substrate. Both parallel (YY=HH) and cross (YX=HV) scattering geometries were used. The measurements were performed starting from $\psi = 0^0$ (corresponding to Z(Y'Y')Z in Porto notations, where Y'≡ [110] of Si substrate) and ended at $\psi = 235^0$. We measured the polarized Raman spectra of several 3PTH samples and one MTHT sample. The results of two rotational-angle dependent Raman measurements are presented in Fig. 10 a and b. We used the $I_D/I_G$ ratio because $I_G$ is independent on the angle ψ. In the measured Raman spectrum of a 3PHT sample in HH scattering geometry, we established that the $I_D/I_G$ ratio moderately increases from $0^0$ to $45^0$, then it significantly drops upon changing of the angle from $45^0$ to $135^0$ and increases again in the interval between $135^0$ and $225^0$, which follows roughly the $I_D \sim \cos^2(\psi)$ law (Fig. 10 a). The intensity ratio was found to vary between 0.6 and 0.85. The polarized Raman spectra, taken in similar conditions from a control sample from the MHTH series with deposited few-layered defected graphene, does not show any simple dependence (Fig. 10 b). We cannot deduce any systematic dependence on the dependence of the angle ψ in HV scattering geometry in both specimens (see HV-marked traces in Figs. 10 a and b).

## 4. Discussion

The above results can explain the relatively high intensity of the D band in the predominantly single-layered graphene of the 3PHT series. Namely, the intensity of the D band decreased to about 60% of the intensity of the G band in absence of graphene edges enhancing the D band appearance. Therefore we can relate the high intensity of the D band to the presence of a large number of defects in graphene films and of micro-sized graphene flakes, bordered by mixed zigzag and armchair shaped edges. It seems that the mix is dominated by armchair shaped edges because the pure zigzag edges do not influence the intensity of D band [27].

However, the observed broadening and blue-shift of the 2D band cannot be clearly explained. We showed in our recent work [28] that the 2D band in exfoliated flakes by the "scotch-tape method" has similar characteristics, namely, the FWHM remains in the range 38-

40 cm$^{-1}$ and is observed at 2658- 2660 cm$^{-1}$, but it is likely that graphene in [28] was not separated from the interlayer of predominantly amorphous carbon. Moreover, the 2D band in a graphene film undoubtedly separated from the interlayer (exfoliated on epoxy resin) has FWHM of 27-29 cm$^{-1}$ and is observed at 2555 cm$^{-1}$. Therefore, we can attribute this broadening of the 2D band to the influence of amorphous and DLC enriched interlayers.

It should be clearly noted that we could not deposit graphene/defected graphene phases directly on the Si substrates.

We did not observe any feature of $C_{60}$ and $C_{70}$ fullerenes by the conducted Raman studies and assumed that this is correlated with the experimental conditions. Namely, we present results of unpolarized and polarized Raman spectra of the surface of the as-grown specimens. On the contrary, we observed the main Raman features of $C_{70}$ fullerenes with decreasing intensity with interlayers' depth in samples after exfoliation by the scotch-type method [28].

According to the above results we can conclude that the deposition takes place in few stages:

-initial formation of stable seeds of a sp$^3$-hybridized carbon phase, mixed with some SiC and SiO$_2$ species, most probably at surface structural defects (dislocations etc.), and a sp$^2$-hybridized carbon phase, mixed with some Si-O species;

-further deposition of thin DLC films in the range of few nanometers on the area, occupied by sp$^3$-dominated carbon phases and of a thin amorphous carbon (aC) films on the area, occupied by sp$^2$-dominated carbon phase;

-deposition of defected graphene on both areas at the final stage of the process.

The synthesis of a separate phase with diamond structure on the surface of the films from MTHT series is unexpected. However, it could be recalled that the deposition of sp$^3$-hybridized carbon films and diamond nano- to micro-crystals was already reported earlier [29]. The MTHT series are deposited at high temperatures and for a relatively long deposition time and we can suppose that the formation of a completely sp$^3$-hybridized carbon phase at suitable positions on the R areas occurs.

**5. Conclusions**

We reported the deposition of few-layered graphene on (001) Si substrates by the CVD process via thermal decomposition of acetone. The deposition is carried out at high temperatures of 1150-1160$^0$C in an Ar main flow. The deposited layers were examined by SEM, XRD, GIXRD, XPS, and Raman spectroscopies. We observed that predominantly single-layered defected graphene (graphene flakes with area of several micrometers) is deposited by use of a pulsed regime, while the deposition for 9 and 12 minutes and similar other conditions resulted in bi- to few-layered defected graphene. We did not perform a successful deposition of graphene/graphene-related phases on Si substrates directly by the thermally-assisted CVD process. We established that the deposition takes place arbitrarily on two types of interlayers: i) the first one consists of a mix sp$^2$-hybridized highly oriented pyrolytic graphite (HOPG) and amorphous carbon, containing also some sp$^3$-hybridized carbon (in the range of a few molar %) as well as small quantity of $C_{70}$ and $C_{60}$ fullerenes; ii) the second one is a dominated by a DLC mix of sp$^3$- and sp$^2$-hybridized carbon, some cubic SiC (in the range of few molar %) as well as small quantity of $C_{60}$ and $C_{70}$ fullerenes. We observed also AA-stacked graphene and some nano-diamond fractions on the top surface of samples, deposited at 9 and 12 min deposition time, respectively. On the contrary, the SiC and fullerenes fractions increase with the depth of the interlayers, allowing to conclude that the initial stages of the CVD process comprise concurrent deposition of SiC and SiO$_2$ species, which determines the formation of the DLC-enriched interlayers as well as deposition of Si-O-C groups that determines the formation of aC-enriched interlayers. Further on, we can confirm that small quantity of $C_{70}$ and $C_{60}$ fullerenes are formed mainly during the initial stages of deposition process, especially, during the experiments using "pulsed" regime of ($C_3H_6O$ +Ar) gas-mix feeding.

## Acknowledgments

The authors gratefully acknowledge support from MPNS COST ACTION MP1204: TERA-MIR Radiation: Materials, Generation, Detection and Applications.

**FIGURES**

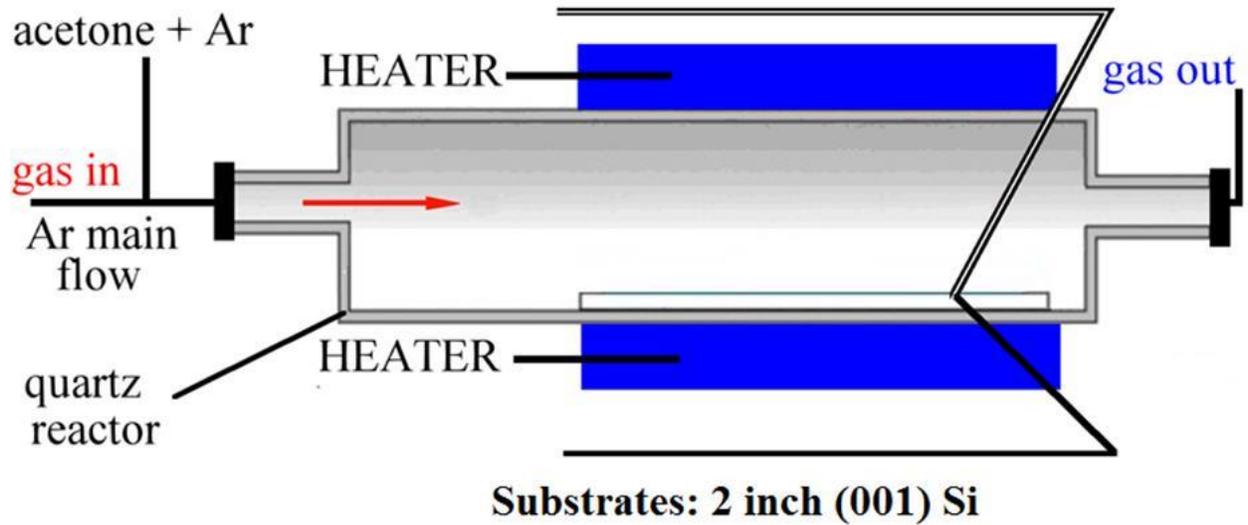

Fig. 1. Schematic of the experimental setup for CVD of graphene/graphene-like films on (001) Si substrates by thermal decomposition of acetone.

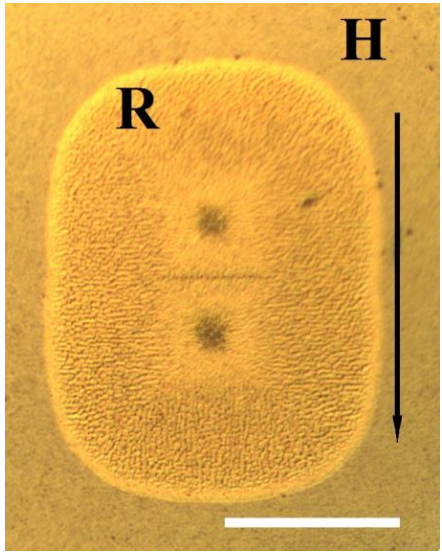

Fig.2. Optical microscopy image of the surface morphology of as-deposited graphene/graphene related phases from the STHT experimental series. The area covered by a clear relief is marked as R and the remaining area is marked as H. The arrow remarks [001] direction of Si substrate and the marker represents 20 μm.

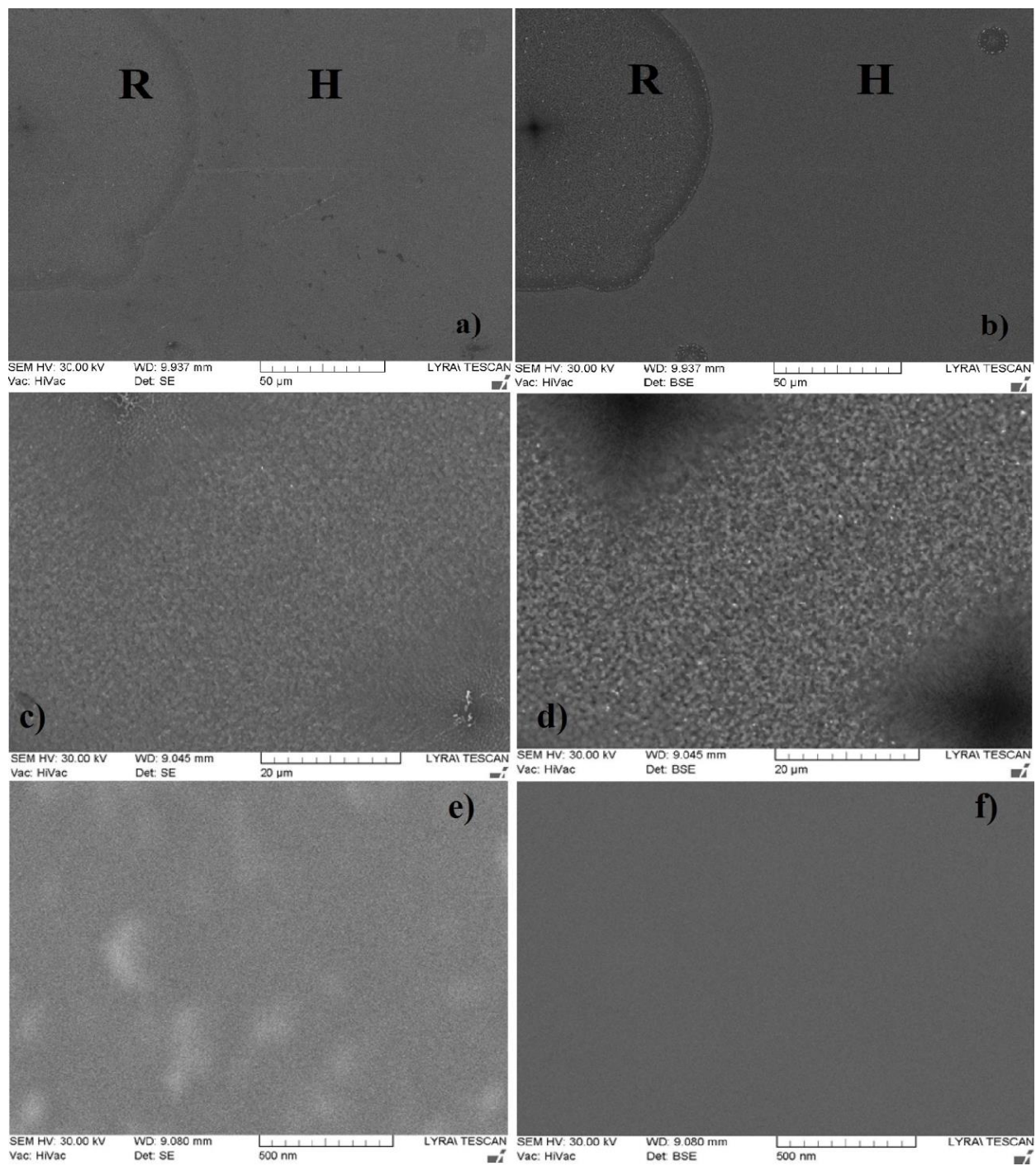

Fig. 3. Secondary electron image (SE) (a) and backscattered electron image (BSE) (b) taken from neighboring R and H areas on the Si substrate (STHT experimental series). Higher resolution SE image (c) and BSE image (d), taken from the R area on the Si substrate. Higher resolution SE image (e) panel and BSE image (f), taken from the H area on the Si substrate.

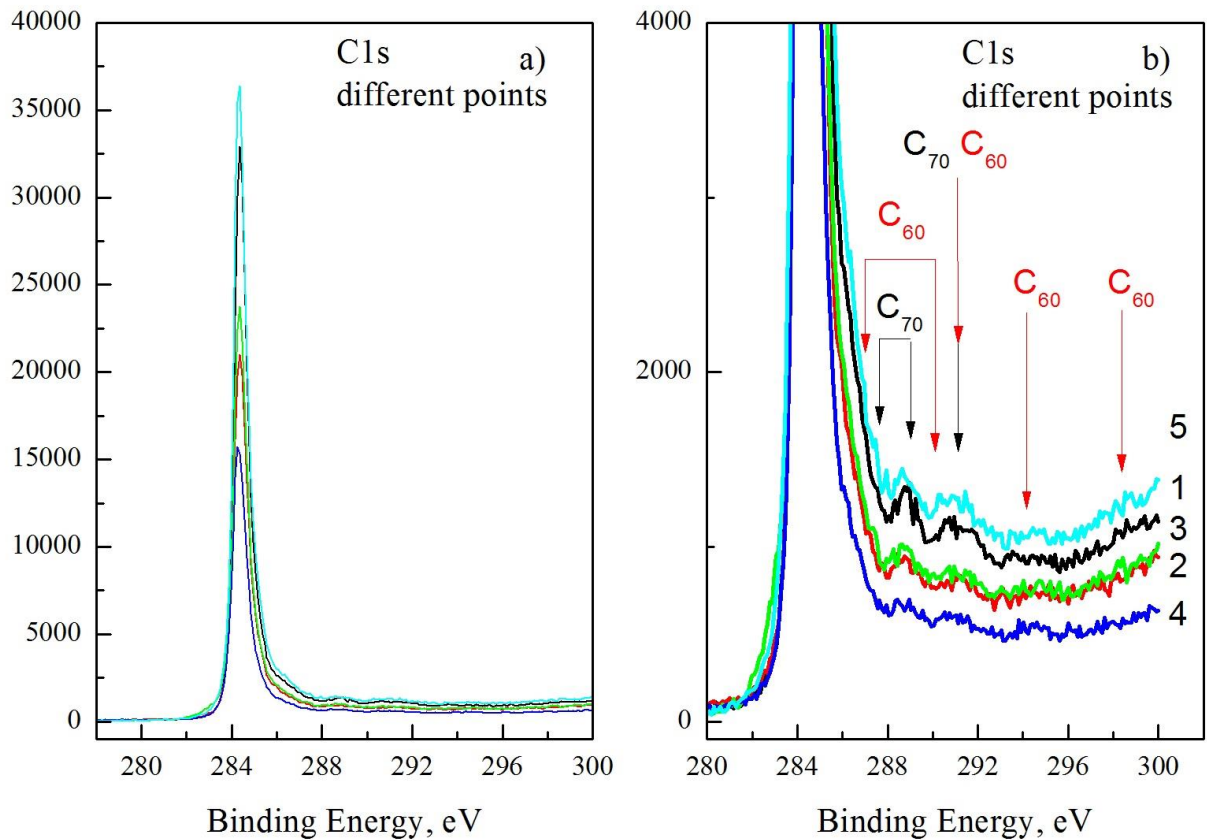

Fig. 4 a) The C1s photoelectron spectra taken from points from 1 to 4. b) Magnified XPS spectra from panel (a) in the area of shake-up satellites of the C1s peak (286-298 eV). The arrows indicate the position of shake-ups of $C_{60}$ and $C_{70}$ fullerenes in the (b) panel.

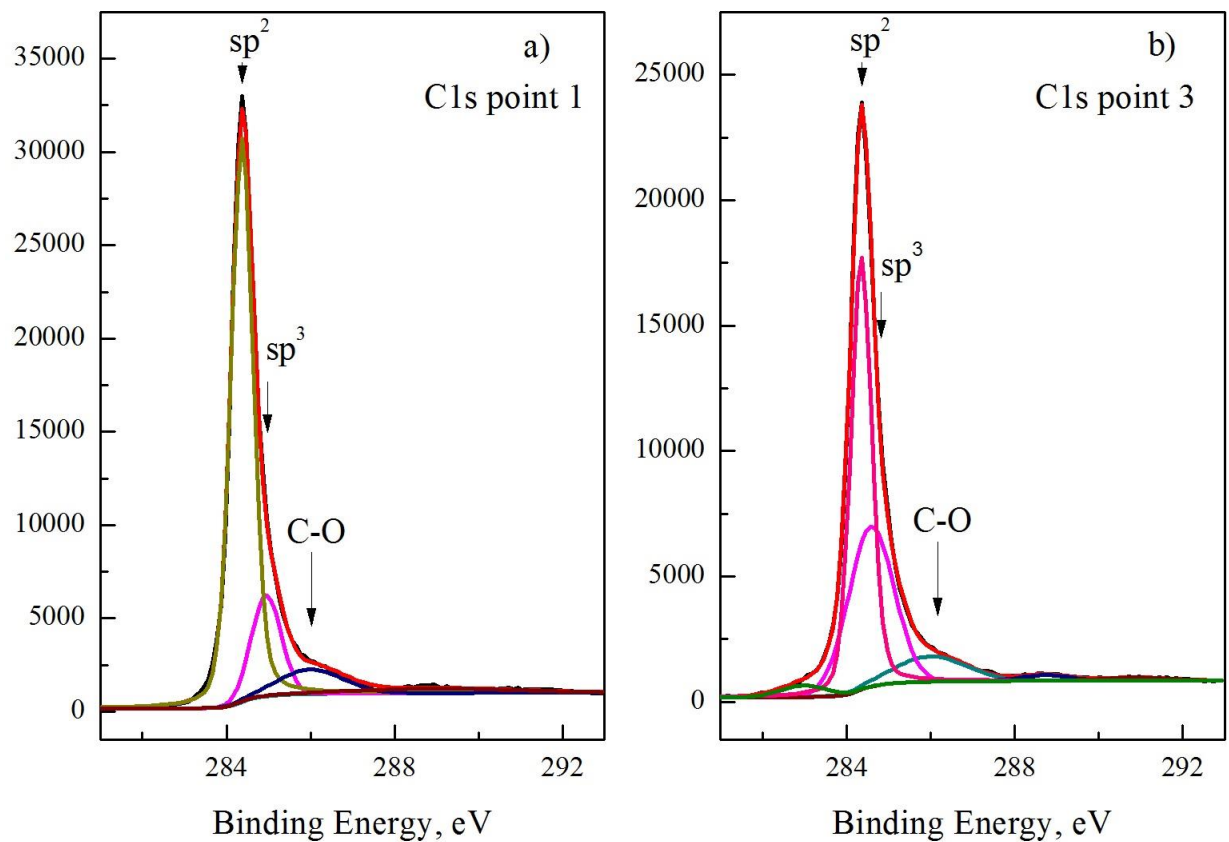

Fig. 5 Deconvoluted XPS C1s lines, taken from point 1 (a) and point 2 (b).

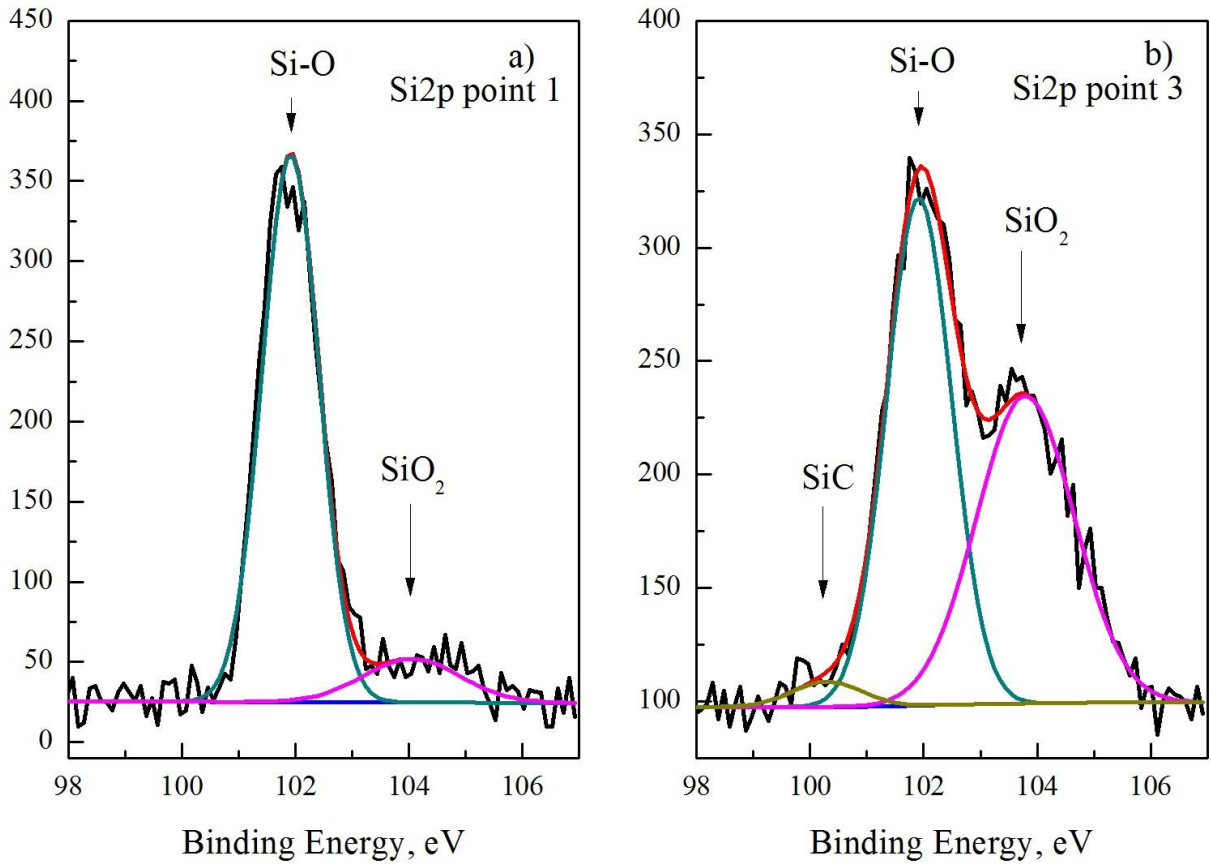

Fig. 6. The Si2p XPS lines measured in point 1 (panel a) and point 2 (panel b).

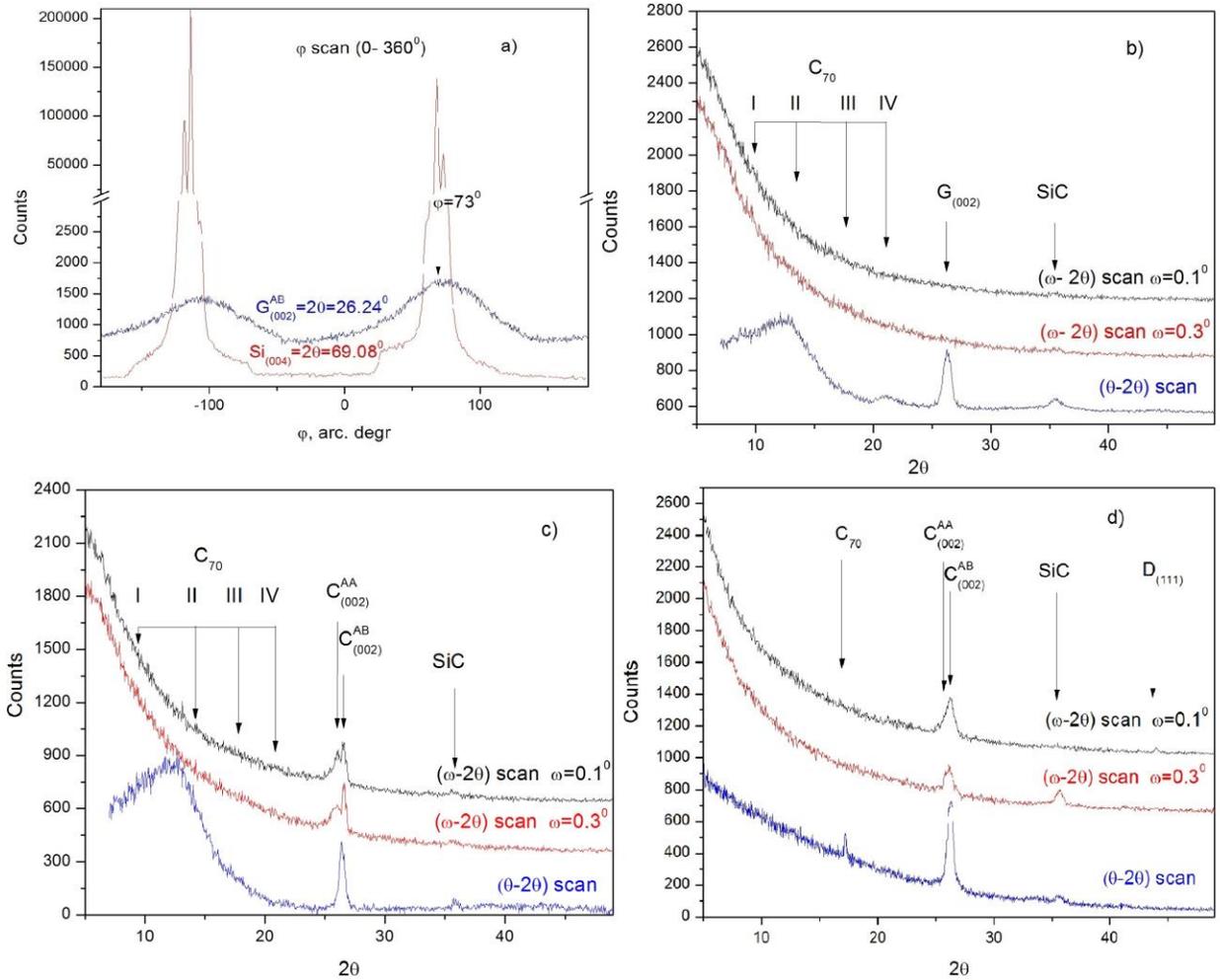

Fig. 7 (a) The XRD patterns of the φ scans in the range φ= 0- 360$^0$ with reference to C$_{(002)}$ and Si$_{(004)}$ reflections, taken from a 10x10 mm$^2$ sample of the MTHT series. (b) The GIXRD and XRD pattern in the (ω-2θ) and (θ- 2θ) scans, taken from a sample from the 3PHT series at ω=0.10$^0$ and ω=0.30$^0$ (black and red traces, respectively). The blue trace represents the XRD pattern from the (θ- 2θ) scan. (c) Results of measurements, similar to those shown in panel (b), but taken from a sample from the STHT series. (d) Results of measurements, similar to those shown in panels (b) and (c), but taken from a sample from the MTHT series.

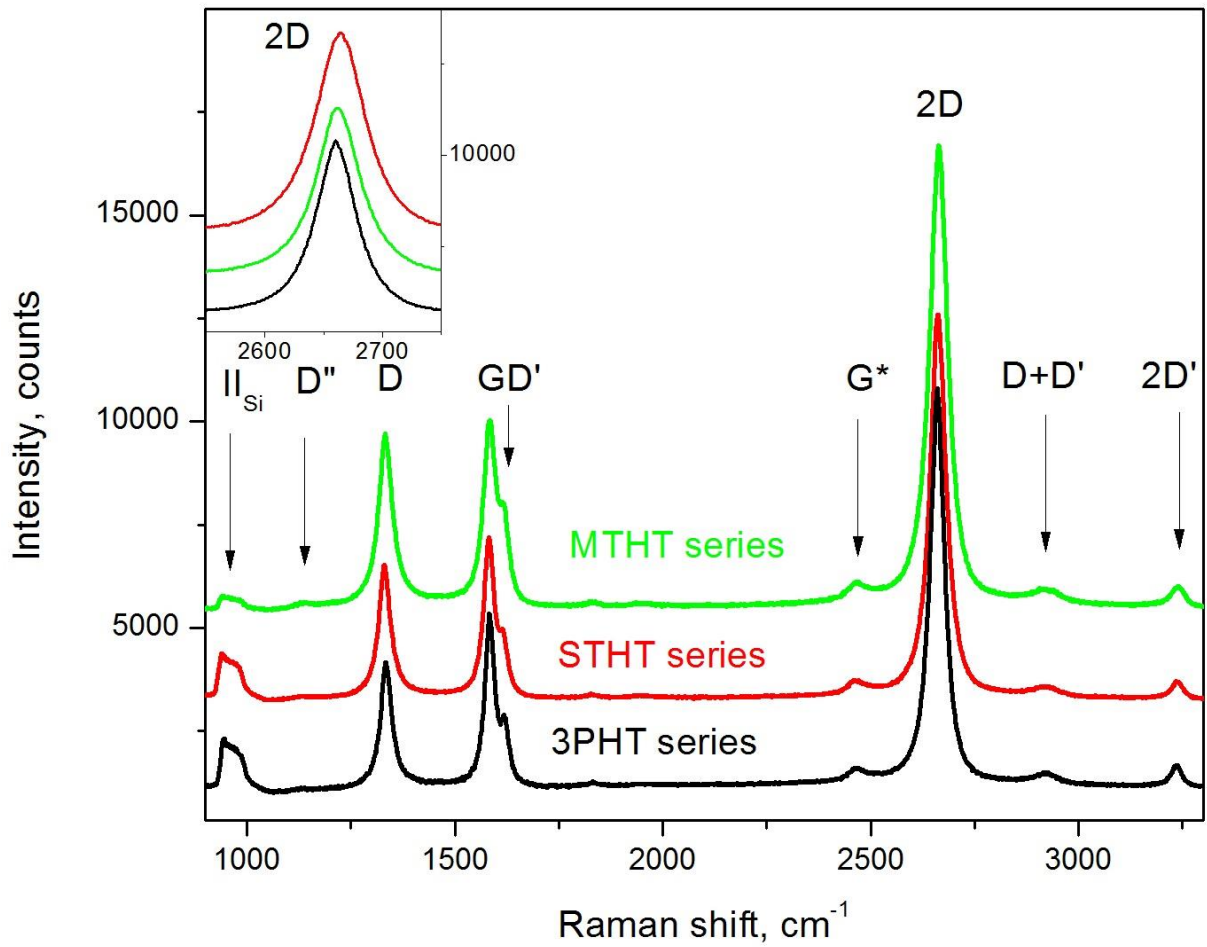

Fig. 8. Typical unpolarized Raman spectra taken from specimens from 3PHT, STHT and MTHT experimental series. The inset shows the magnified part around the 2D band.

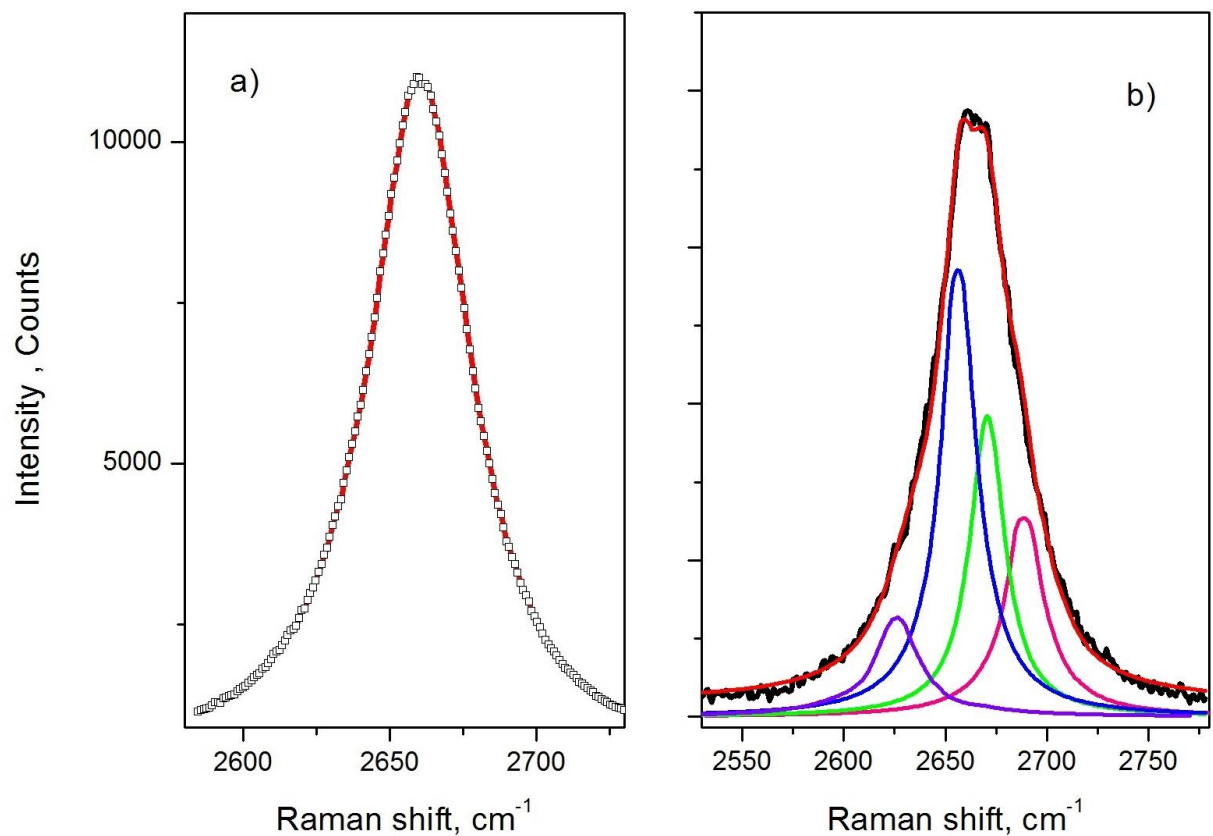

Fig. 9. Deconvolution of the 2D band, identified as coming from single layered 3PHT series (a) and bi- layered STHT series (b). The empty black squares and the black line on panels (a) and (b) are the as-measured signal, while the red lines show the fitted Lorentzians.

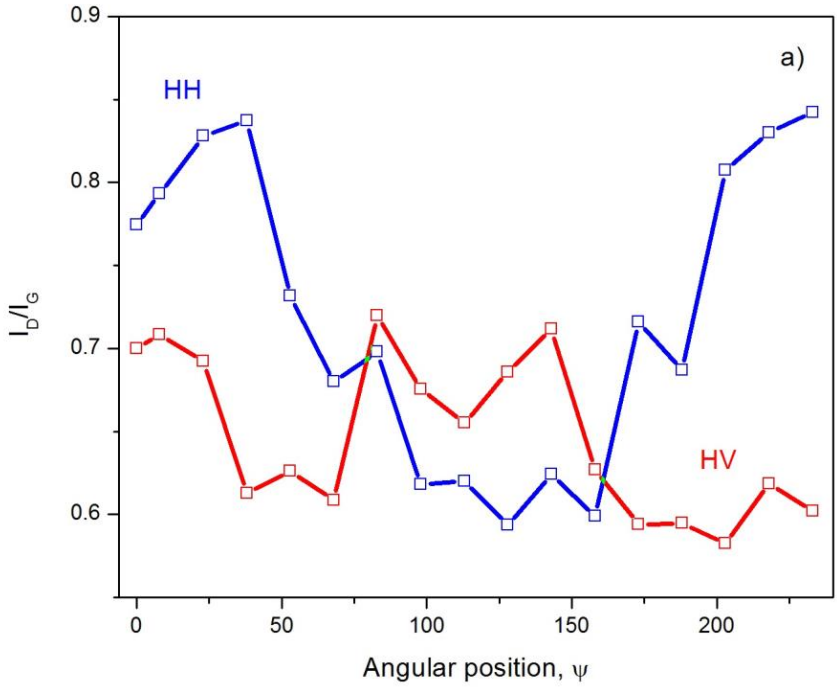

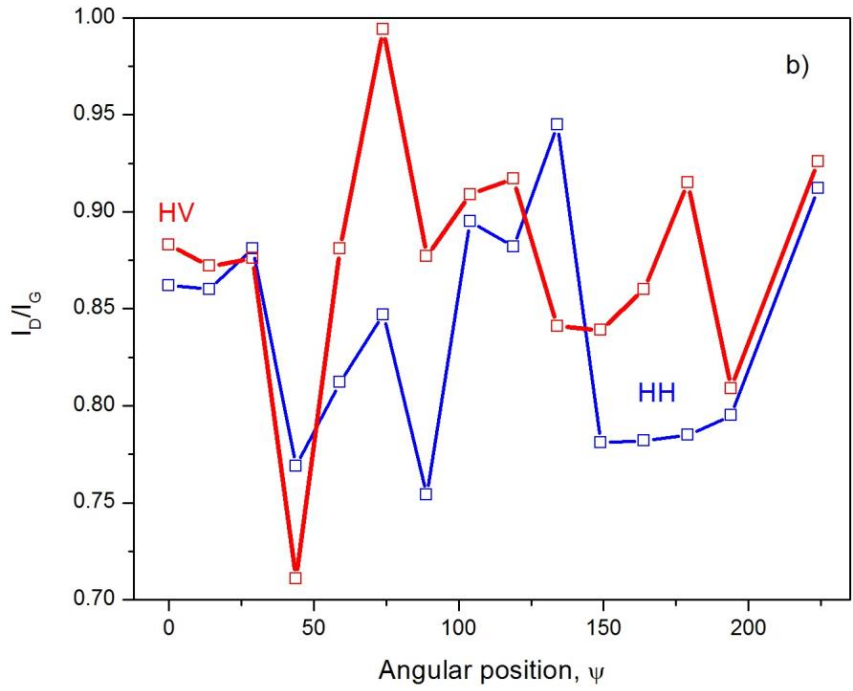

Fig. 10 Polarized Raman spectra taken from 3PHT series (panel a) and MTHT series (panel b).

**TABLES**

| Process designation | Ar / (Ar+C$_3$H$_6$O) gas flows ratio, ccm/min | Pulsed regime/ number of pulses | Deposition temperature,$^0$C | Deposition time, min |
|---|---|---|---|---|
| **STHT** | 180/30 ccm/min | No | 1150-1170 | 9 |
| **3PHT** | 180/30 ccm/min | Yes/ 2 pulses | 1150-1170 | 9 |
| **MTHT** | 180/30 ccm/min | No | 1150-1170 | 12 |

Table 1. Summarized parameters of the deposition processes.

| point | C, at% | O, at. % | Si, at. % | O/C | Sp$^3$/sp$^2$ | % Si in SiC | % Si in Si-O | % Si in SiO$_2$ |
|---|---|---|---|---|---|---|---|---|
| 1 | 95.62 | 3.17 | 1.21 | 0.03 | 0.21 | - | 88.1 | 11.9 |
| 2 | 90.32 | 7.73 | 1.95 | 0.08 | 0.73 | 2.7 | 51 | 46.3 |
| 3 | 94.64 | 4.57 | 0.79 | 0.04 | 0.60 | - | 92 | 2 |
| 4 | 94.89 | 4.01 | 1.10 | 0.04 | 0.70 | ~ 1 | 87.4 | 12.6 |

Table 2. Detailed results of the XPS measurement on a sample from the STHS series. The spectra are taken from five neighboring points on a linear segment, separated at about 0.5 mm, in each one of the measured samples.

| Experimental series | D band, cm$^{-1}$ | G band, cm$^{-1}$ | 2D band, cm$^{-1}$ | FWHM 2D Band, cm$^{-1}$; assignment |
|---|---|---|---|---|
| 3PHT | 1334 | 1581 | 2660 | 40-42; SL |
|  |  | 1582 | 2662-2664 | 48-54; BL |
| STHT | 1331 | 1582 | 2662-2664 | 48-54; BL |
|  | 1331 | 1583 | 2664-2668 | >54 ML |
| MTHT | 1333 | 1583 | 2664-2668 | >54 ML |

Table 3. Summary of the results of Raman measurements of as-deposited defected graphene films from different experimental series. The abbreviations SL, BL and ML denote single-, bi- and multi-layered graphene.